\documentclass[aps,prl,onecolumn,notitlepage,showpacs]{revtex4-2}

\usepackage{bm}
\usepackage{graphicx}
\usepackage{latexsym}
\usepackage{amsmath}
\usepackage[colorlinks=true,citecolor=blue,linkcolor=blue]{hyperref}
\usepackage{bbding}
\usepackage[caption=false]{subfig} 
\usepackage{float}
\usepackage{subfloat}
\usepackage[section]{placeins}
\usepackage{csquotes}
\usepackage{relsize}
\usepackage{enumitem}
\usepackage[Symbol]{upgreek}
\DeclareGraphicsExtensions{.pdf,.png,.jpg}

\LetLtxMacro{\OldSqrt}{\sqrt}
\newcommand{\ClosedSqrt}[1][\hphantom{3}]{\def\DHLindex{#1}\mathpalette\DHLhksqrt}
\makeatletter
    \newcommand*\bold@name{bold}
    \def\DHLhksqrt#1#2{%
        \setbox0=\hbox{$#1\OldSqrt{#2\,}$}\dimen0=\ht0\relax%
        \advance\dimen0-0.2\ht0\relax
        \setbox2=\hbox{\vrule height\ht0 depth -\dimen0}%
        {\hbox{$#1\expandafter\OldSqrt\expandafter[\DHLindex]{#2\,}$}
        \lower\ifx\math@version\bold@name0.6pt\else0.4pt\fi\box2}
    }
    \renewcommand*{\sqrt}[2][\ ]{\ClosedSqrt[\leftroot{-2}\uproot{1}#1]{#2}\kern0.1em} 
\makeatother

\begin{document}

\title{Solid-state Tube Wakefield Accelerator \\ using Surface Waves in Crystals}

\author{Aakash A. Sahai}
\address{College of Engineering and Applied Science\\ University of Colorado, Denver, CO 80204\\ aakash.sahai@gmail.com}

\author{Toshiki Tajima, Peter Taborek}
\address{Department of Physics \& Astronomy and Applied Physics\\ University of California, Irvine, CA 92697}

\author{Vladimir D. Shiltsev}
\address{Accelerator Research Department\\ Fermi National Accelerator Laboratory, Batavia, IL 60510}

\begin{abstract}
Solid-state or crystal acceleration has for long been regarded as an attractive frontier in advanced particle acceleration. However, experimental investigations of solid-state acceleration mechanisms which offer $\rm TVm^{-1}$ acceleration gradients have been hampered by several technological constraints. The primary constraint has been the unavailability of attosecond particle or photon sources suitable for excitation of collective modes in bulk crystals. Secondly, there are significant difficulties with direct high-intensity irradiation of bulk solids, such as beam instabilities due to crystal imperfections and collisions etc. 

Recent advances in ultrafast technology with the advent of submicron long electron bunches and thin-film compressed attosecond x-ray pulses have now made accessible ultrafast sources that are nearly the same order of magnitude in dimensions and energy density as the scales of collective electron oscillations in crystals. Moreover, nanotechnology enabled growth of crystal tube structures not only mitigates the direct high-intensity irradiation of materials, with the most intense part of the ultrafast source propagating within the tube but also enables a high degree of control over the crystal properties.

In this work, we model an experimentally practicable solid-state acceleration mechanism using collective electron oscillations in crystals that sustain propagating surface waves. These surface waves are driven in the wake of a submicron long particle beam, ideally also of submicron transverse dimensions, in tube shaped nanostructured crystals with tube wall densities, $n_{\rm tube}\sim10^{22-24}\rm cm^{-3}$. Particle-In-Cell (PIC) simulations carried out under experimental constraints demonstrate the possibility of accessing average acceleration gradients of several $\rm TVm^{-1}$ using the solid-state tube wakefield acceleration regime. Furthermore, our modeling demonstrates the possibility that as the surface oscillations and resultantly the surface wave transitions into a nonlinear or ``crunch-in'' regime under $n_{\rm beam}/n_{\rm tube} \gtrsim 0.05$, not only does the average gradient increase but strong transverse focusing fields extend down to the tube axis. This work thus demonstrates the near-term experimental realizability of Solid-State Tube Wakefield Accelerator (SOTWA).

The ongoing progress in nanoengineering and attosecond source technology thereby now offers the potential to experimentally realize the promise of solid-state or crystal acceleration, opening up unprecedented pathways in miniaturization of accelerators. 
\end{abstract}

\maketitle

\section{Introduction}

Particle acceleration techniques using collective charge density oscillations in crystals have been known to be an attractive possibility for the past many decades \cite{atomic-accelerator}.

\vspace{-3.5mm}
\subsection{Solid-state Acceleration using Wakefields in Crystal Plasmas: \\ Attosecond sources and Crystal tubes}\label{sec:lpa-intro}

In solid-state or crystal acceleration mechanisms a charged particle beam gains energy by extracting the electromagnetic field energy of collective electron oscillation modes excited in crystals. These solid-state collective oscillations are known to sustain propagating charge density waves of high energy densities. These collective oscillations and the associated waves can be efficiently excited as wakes of pulsed sources of particles or photons with pulse dimensions that are resonant with the scales of collective oscillations in solid-state. However, the theoretically modeled solid-state acceleration gradients \cite{Tajima-crystal-xray, Chen-solid-state} which are known to be orders of magnitude higher than the time-tested radio-frequency technology as well as the emerging gaseous plasma acceleration \cite{Tajima-Dawson-Laser-wakefield,Chen-Dawson-Beam-wakefield} techniques, are yet to be experimentally verified and further studied.  

Experimental verification of solid-state acceleration mechanisms has been so far hampered by several technological challenges such as unavailability of pulsed particle and photon sources that are resonant with the collective oscillations in crystals. However, technological advances in intense particle and photon pulsed ultrafast source compression technologies have continued to drive the pulse dimensions towards ever shorter time and spatial scales. These technological advancements in ultrafast source compression techniques have made scales required to resonantly excite collective electron modes for solid-state acceleration mechanisms experimentally accessible. Especially, recent breakthroughs in attosecond scale photon \cite{Atto-xray-pulse} and particle \cite{Atto-particle-bunch} bunch ultrafast source technologies have opened up the potential for experimental realization of long-sought solid state acceleration \cite{atomic-accelerator,Tajima-crystal-xray, Chen-solid-state}. 

Although attosecond source technologies provide an effective means for resonant excitation of collective modes in solid-state crystal media, there still exist other technological barriers. In addition to the barriers due to the scarcity of attosecond sources, accessing solid-state gradients has also been impeded by difficulties with direct irradiation of solids at high intensities using particle or photon beam. Advances in nano-structured materials and nanoengineering of tube-like structures in crystals, such as nanotubes, however now offer the possibility of overcoming these difficulties with direct interaction of a crystal with high-intensity sources. Direct interaction of a high-intensity particle beam is reported to undergo severe filamentation due to the deformities in the crystal structure \cite{beam-crystal-filamentation}. Not only is a filamented beam detrimental to driving a coherent wake but it also leads to a severely uncontrolled interaction and energy dissipation.

Solid-state plasmas with electron densities $n_0 \sim 10^{22-24}{\rm cm^{-3}}$, sustain electron oscillations at superoptical time, ${\rm 177 (} n_0 {\rm [10^{22} cm^{-3}] )^{-1/2}}$ attosec ($\sim\omega_{pe}^{*-1}$ where $\omega_{pe}=(n_0e^2\epsilon_0^{-1}m^{-1})^{1/2}$and $m_e$ the electron mass \cite{Bohm-Pines-electron-gas}) and spatial scales, ${\rm 330 (}n_0 {\rm [10^{22} cm^{-3}])^{-1/2} ~ nm}$ ($\sim\lambda_{\rm pe}$). Electron modes at such scales offer Tajima-Dawson (wavebreaking) acceleration gradients \cite{Tajima-Dawson-Laser-wakefield} of the order of, $E_{wb} \simeq 9.6 ( n_0 [10^{22} cm^{-3}] )^{-1/2} ~ \rm TVm^{-1}$. By coupling with these superoptical scales, submicron particle bunches (e.g., planned $\sigma_z<1\mu m$ \cite{yakimenko-facet-ii}) or intense keV photon lasers \cite{xray-wakefield-2016} make excitation of unprecedented $\rm TVm^{-1}$ average gradients experimentally feasible. 

\vspace{-2.0mm}
\subsection{Progress of Wakefields Acceleration in Gaseous Plasmas: \\ Femtosecond sources}\label{sec:gas-plasma-acc}

Over the past few decades, access to tens of femtosecond chirped pulse amplified \cite{morou-CPA} $0.8\mu m$ wavelength lasers (with few femtosec single cycle) and particle bunches has enabled experimental verification of advanced particle acceleration techniques that use collective electron oscillations in gaseous plasmas \cite{Tajima-Dawson-Laser-wakefield,Chen-Dawson-Beam-wakefield}. Whereas lasers have been compressed to few cycle long pulses using the innovative chirped pulse amplification technique \cite{morou-CPA}, ultrashort particle bunches have been obtained via phase-space gymnastics \cite{phase-space-gym} or self-modulation in plasma \cite{proton-Beam-Wakefield-Expt}. Both these ultrafast source technologies have enabled successful gaseous plasma acceleration experiments with many $\rm GVm^{-1}$ gradients \cite{CPA-Laser-Wakefield-Expt,elec-Beam-Wakefield-Expt}. These experiments have used micron-scale charge-density waves in gaseous bulk plasma. 

Control over bulk plasma waves in homogeneous gases by femtosecond-scale sources \cite{Tajima-Dawson-Laser-wakefield,Chen-Dawson-Beam-wakefield} has lead to the successful demonstration of gaseous plasma wakefield acceleration techniques. Numerous advantages of these techniques over conventional radio-frequency acceleration techniques has now lead to them being enhanced and fine-tuned for real-world applications using commercially available femtosecond sources. Some of these enhancements include control over: (a) wakefield profile distortions from ion motion \cite{ion-motion-intense-beam}, (b) dark current injection and acceleration due to secondary ionization \cite{dark-current}, (c) accelerated beam emittance growth due to scattering off of plasma ions \cite{ion-motion-beam-emittance}, (d) positron defocusing by the bared ions, (e) repetition rate constraints due energy coupling to long-lived ion modes \cite{ion-wake-excitation} etc.

Non-homogeneous plasmas of specific shapes have been proposed to address many of the above enhancements of gaseous bulk plasma acceleration. The earliest shaped plasma proposal \cite{Tajima-Plasma-Fiber-Acc} for a fiber accelerator sought to keep high-intensity laser pulses continuously focussed \cite{Tajima-Laser-Hollow-Guide}. Utilizing this shaped plasma proposal, gaseous plasma fibers that are excited using mechanisms such as laser-heated capillary \cite{laser-heated-capillary} etc. are now regularly used for plasma fiber guided laser-plasma acceleration. Mechanism of beam-driven shaped gaseous hollow plasma (later labelled hollow-channel) acceleration has also been studied \cite{Katsouleas-PRL-1998}. Experiments on beam-driven gaseous hollow plasma using intense positron beams have observed $\rm\sim 200 MVm^{-1}$ peak gradients ($\sim0.01E_{wb}$) \cite{FACET-hollow-2016}. Access to higher gradients in shaped gaseous hollow plasmas is currently under active research. Active areas of research include technological difficulties in shaping a desired channel in gaseous plasmas apart from challenges due to the absence of any focusing force \cite{Katsouleas-PRL-1998} such as control of higher-order transverse wakes excited by the drive beam due to its misalignment from channel axis \cite{higher-order-transverse-wakes} and beam-breakup resulting from these transverse wakes. Recent results have demonstrated that beam breakup may be controllable via further shaping of the gaseous hollow plasmas \cite{pukhov-axial-spike-channel}.

In this paper, we introduce and model a regime of experimentally realizable solid-state acceleration that uses charge density waves of submicron scale lengths in nanotube shaped solid-state plasmas. The acceleration modes in this regime of solid-state tube wakefield acceleration take advantage of the developments in nano-fabrication as well as submicron particle or attosecond photon pulsed source technology. We show using analytical and computationally modeling that the crystal tube surface electron oscillations sustain an electrostatic ``crunch-in'' mode \cite{xray-wakefield-2016,Sahai-PhD-thesis,ion-wake-excitation}. This electrostatic mode supports electromagnetic surface wave modes with phase velocity close to the driver velocity and on-axis longitudinal electric fields that approach the Tajima-Dawson gradient of the tube wall electron density. This high phase velocity surface wave mode supported by excitation of tube wall electron oscillations makes solid-state tube wakefield accelerator regime quite effective. 

Although significantly different from traveling wave modes supported by electron oscillations in solid-state, a similar electron oscillation mode of gaseous plasma hollow channels has been computationally observed in a few previous works. However, neither its structural and electromagnetic properties nor its acceleration characteristics have been extensively modeled. In gaseous plasmas the ``crunch-in" like mode has been observed in simulation works that have used experimentally feasible parameter regime such as a laser-driven shaping of a hollow plasma proposal \cite{Kimura-2011}, a proton beam driven shaped hollow plasma acceleration proposal in externally magnetized plasma \cite{Li-2015} and a electron or positron beam driven shaped hollow plasma \cite{ion-wake-excitation}.

An important recent work \cite{xray-wakefield-2016, Sahel-PoP-2018} has recently studied and modeled the excitation of modes in crystal tubes using attosecond keV photon x-ray pulses. This work on modeling of x-ray wakefield tube accelerator has demonstrated the potential of using x-ray wakefield acceleration mechanism in tubes for sustaining many $\rm TV\text{-}cm^{-1}$ gradients. With the advent of a few cycle high-intensity x-ray laser using thin-film compression technique, the crystal x-ray wakefield acceleration mechanism has the potential to further advance the progress made by the Ti:Sapphire 800nm optical laser based gaseous plasma wakefield acceleration technique.

However, the mechanism of beam-driven surface modes in bulk crystals and crystal tubes, as opposed to those driven by an x-ray laser, has not yet been modeled and characterized. This is especially important due to the recent opening up of the availability of submicron particle bunches. The beam-driven crystal tube phenomena investigated and the results reported here indicate that the x-ray driven crystal tube wakefields characterized in \cite{xray-wakefield-2016, Sahel-PoP-2018, Sahel-XTALS-2019} are quite similar to that in the beam-driven crystal tube case. Therefore, our work shows that crystal tube wakefields have both longitudinal and focusing fields similar to the x-ray driven wakefields \cite{xray-wakefield-2016, Sahel-PoP-2018, Sahel-XTALS-2019}. It may be noted that our work on beam-driven wakefields in a crystal tube is distinctive from previously modeled gaseous hollow-plasma wakefields because in gaseous hollow-plasma the wakefields of a relativistic particle beam are proven to have zero focusing fields \cite{zero-focussing-field}. The preliminary analysis and computational results presented below demonstrate the experimental realizability of Solid-State Tube Wakefield Accelerator (SOTWA).

In the following sections on modeling of beam-driven wakefields in crystal tubes, we introduce and characterize the beam-driven solid-state tube accelerator using surface wave wakefields in crystals. The model and significance of solid-state collective electron or plasmon oscillation modes is presented in sec.\ref{sec:quant-plasmon}. An analytical model of the tube wall electron oscillations extending into the tube is presented in sec.\ref{sec:surface-oscillation-theory}. Preliminary proof-of-principle particle-in-cell method based computational modeling of beam-driven solid-state tube wakefield accelerator is detailed in sec.\ref{sec:crunch-in-PIC-simulations}. We also study the novel ``crunch-in'' behavior shown by the wakefields in a tube which includes wakefield amplitudes close to the Tajima-Dawson acceleration gradient for relatively small beam to tube density ratios as well as the existence of transverse fields that extend down to the tube axis.

\section{Collective oscillation in quantum mechanical systems: \\
oscillation modes of electron gas in Crystal ionic lattice}\label{sec:quant-plasmon}

Collective electron oscillations in crystals have for long been established a critical yet physically valid simplification of the many body interaction in solid-state materials. The many body problem of solid-state electrons can be either described using an assembly of Fermions or using collective oscillation theory. The collective oscillation approach was exhaustively modeled in theory \cite{Bloch-1929, Tomonaga-1950, Pines-Bohm-1, Pines-Bohm-2} (phonon, plasmon and polaritons) and experimentally proven to result in observable effects \cite{Ritchie-surface} in 1950s.

Solid-state collective oscillations were first investigated with great details in the context of the modeling of the stopping power of an incident electron beam in metals with an inherent crystal structure \cite{Pines-RMP-1956}. The predictions of energy loss of a particle beam incident on a metal were found to be in excellent agreement with the theory of excitation of collective electron oscillations in the crystal, driven as a wake of the incident particles.

The terminology of excitation of collective oscillations in the \enquote{wake} of an incident particle was introduced in 1950s. Moreover, to explain quantization in beam energy loss when interacting with a thin foil with thickness of the order of mean free path of the bulk plasma oscillations in crystals, these oscillations where referred to as plasmons. The theoretical plasmon model of the collective oscillations of electrons in crystals showed good agreement with experiments on energy loss of injected beam electrons. In addition to the explanation of the quantized energy loss of the beam, the conditions for the excitation of collective oscillations in the wake of an incident particle were also detailed. The collective oscillations of valence electrons were demonstrated to be quite similar to the plasma oscillations observed in gaseous plasmas.

Bloch \cite{Bloch-1929} was the first to model the excitations of a Fermi gas as collective gas oscillations as opposed to excited states of single particles. Bloch treated the Fermi gas collective oscillations both with and without quantum mechanics. However, when density fluctuations were important to be considered for understanding the phenomena, then quantum aspects of the problem were critical.  In Tomonaga's work \cite{Tomonaga-1950} on collective oscillations it was demonstrated that modeling and understanding the many Fermion interaction in solid-state electron gas in a crystal was greatly simplified by the use of collective modes of many Fermion oscillations. These collective electron oscillations were first investigated in 1D in 1950 by Tomonaga through the use of density fluctuation method (where density if the field variable) as opposed to the conventional quantum mechanical method of computation of expectation values from the wave functions of the system. This was because the equations of motion of collective oscillations are linear in field variable (density fluctuation) as opposed to bilinear field variable terms in the conventional method. Moreover, linearity of the field variables in the equation of motion holds irrespective of the presence or absence of inter-particle forces (direct interaction between single particles).

Subsequently, in 1953 Pines and Bohm \cite{Pines-Bohm-1} used a collective canonical transformation method to analyze the collective many Fermion (electron) oscillations in crystal lattices in metals. They recognized the dominance of the long-range nature of the Coulomb forces which controls the phenomena and produces collective oscillations of clouds of electrons over spatial scale much greater than Debye length. Here the characteristic dimension of an electron cloud is of the order of a Debye length (in quantum mechanical treatment this characteristic dimension is modified).


The collective behavior is therefore critical to explain physical phenomena over micron or nanometer scales. In their work the term ``plasmon'' was introduced to describe the quantum of elementary electron excitation associated with this high-frequency collective motion in bulk crystals with the dimensions of the order of one plasmon oscillation wavelength. This is a quantum of energy of collective oscillations of valence electrons. The energy of a plasmon was shown to be \cite{Pines-Bohm-1, Pines-Bohm-2},
\begin{equation}
\hbar \omega_{pe} = \hbar \left(\frac{4\pi n_0e^2}{m_e}\right)
\end{equation}
When the dimension of the solid-state material is below mean free path of the collective electron oscillations, quantization of electron plasma frequency is observed. Plasmon energy is greater than the energy of any individual conduction band electron. Although, these plasmonic oscillations are the quantum analog of the collective oscillations of plasma electrons in gaseous plasmas their extremely high energies ($\hbar \omega_{pe} \gg k_BT_e$) and small spatial scales necessitate the consideration of the quantum nature of these oscillations. Typical, valence electron density in crystals which lies in the range of $n_0 \simeq \rm 10^{22}-10^{24} cm^{-3}$ result in plasma energies in crystals of $\hbar \omega_{pe} \simeq \rm 4 ~ to ~ 30 eV$. As a result of this, plasmonic collective oscillation are not thermally excitable and under normal conditions metals do not sustain plasmonic oscillations driven by a valence electron.

The dispersion relation of a plasmon using the Hamiltonian approach of a Fermi electron gas in the presence of an ionic lattice was derived in \cite{Pines-Bohm-1, Pines-Bohm-2}. This approach is essential when a quantum-mechanical treatment of the electronic motion is required, as is the case for the electrons in a metal. The particle based or density fluctuation approach taken in this work was argued to be a microscopic approach to the modeling of collective oscillations. In the density fluctuation method, the Coulomb interaction was effectively split up into a long-range and a short-range part. The conditions under which an externally injected electron beam can excite collective electron oscillations in a crystal.

In contrast certain other plasmon models simply used the dielectric constant of a medium to represent its plasma behavior, which is a macroscopic approach. To model the effect of electron-electron interaction on the stopping power of a metal for high-energy charged particles, the electron gas is described as a classical fluid with an artificially introduced coefficient of internal friction (Kronig and Korringa). 

As the collective behavior of the electron gas is essential to model phenomena over distances greater than the Debye length (or a critical spatial scale, quantum mechanically), cumulative potential of all the electrons involved in the oscillation is quite large since the long range of the Coulomb interaction permits a very large number of electrons to contribute to the potential at a given point. The higher the density the larger is the number of electron that contribute to the potential and thus higher is the collective field and potential.

Surface wave modes in solid-state plasmas at the interface of crystals with vacuum or metals have also been well modeled \cite{Stern-Ferrel-1960}. Using both microscopic as well as macroscopic modeling, the ``surface plasmon'' oscillation frequency of a metal vacuum interface is determined to be $\omega^s_p = \omega_{pe}/\sqrt{1 + \epsilon}$, where $\epsilon$ is the dielectric constant of the metal. The dispersion characteristics of surface plasmon and phonon modes have been well characterized in a linearized perturbative regime \cite{Tajima-Ushioda-1978}.

\section{Solid-state Surface Waves in Crystal Tube:\\
surface electron oscillation model in tube nanostructure}
\label{sec:surface-oscillation-theory}

Using the collective electron oscillation models described above and under the condition that the dominant behavior of solid-state media is that of an ideal electron gas, we analytically model surface electron oscillations in a tube structure driven in the wake of an electron beam. These analytically modeled surface electron oscillations also sustain a propagating surface wave which propagates at nearly the same velocity as the drive beam. Moreover, as collision-less behavior dominates it is possible to treat the density fluctuations using a single particle oscillation model. 

Because the crystal tube under consideration here conforms with a cylindrical geometry, in our analysis we model the surface electron oscillations in a cylindrical coordinate system. Moreover, as these surface oscillations and the surface wave sustained by these collective oscillations co-propagate with the electron beam, the longitudinal dimension of the cylindrical coordinates will be transformed to a co-moving frame behind the drive beam. An preliminary analysis of a similar nature has been previously attempted \cite{Sahai-crunch-in}.

Classifying the onset of non-linearity and wavelength of the density oscillations both require understanding of the electron dynamics within the plasma. The plasma considered is one of density $0$ for $r<r_{\rm tube}$ and $n_0$ for $r>r_{\rm tube}$. An electron or positron driving beam of density $n_b$ and volume $\mathcal{V}_b$ -- moving with velocity $v_b \mathbf{\hat{z}} = c\beta_b \mathbf{\hat{z}}$ -- perturbs electrons within the plasma electrostatically. Subsequently, the plasma electrons oscillate freely in the radial direction as a result of the electric field that has been set up due to the no longer quasineutral plasma.

This electric field is found using Gauss' law
\begin{equation}
\int_S \mathbf{E} \cdot d\mathbf{S} = 4\pi ~ Q_{enc}
\label{eq:GaussInt}
\end{equation}
where $S$ is a closed Gaussian surface, $\mathbf{E}$ is the electric field, $d\mathbf{S}$ is the surface area element of $S$, $Q_{enc}$ is the total charge enclosed within $S$, and $\epsilon_0 = 1/(4\pi)$ is the permittivity of free space (in cgs units). As the plasma is cylindrically symmetric, a cylinder of radius $r$ and length $l$ is used as the Gaussian surface $S$. $d\mathbf{S}$ can therefore be simplified to $d\mathbf{S} = r\,d\theta\,dz \mathbf{\hat{r}}$. Assuming the electric field to be purely radial, $\mathbf{E} = E_r\mathbf{\hat{r}}$, the left hand side of Gauss' law is simplified (after integrating) to $2\pi r l E_r$.

The enclosed charge is given by the integral of the ion charge density over the volume enclosed by $S$ (cylinder of radius, r). The ions within the plasma are of density $n_0$. As the plasma density $n_0$ is constant, the enclosed charge is given by the net volume of plasma within $S$ multiplied by $en_0$, $Q_{enc} = en_0\pi (r^2 - r_{\rm tube}^2) l$. Gauss' law thus gives the following form for the radial electric field set up by the non-quasineutral plasma
\begin{equation}
E_r(r) = 4\pi en_0 ~ \frac{1}{2r}(r^2 - r_{\rm tube}^2)
\label{eq:RadEField}
\end{equation}
As the electric field vanishes for $r=r_{\rm tube}$, equation \eqref{eq:RadEField} describes the electric field for an electron situated initially on the channel wall.

The force experienced by a given plasma electron is found by multiplying the electric field by $-e$, the electronic charge. Finally, a transformation to the frame of the driving beam $\xi = \beta_b ct - z$ is made. This is to allow for direct comparisons to be made between the model and Particle-In-Cell simulations (section \ref{sec:crunch-in-PIC-simulations}). The equation of motion is
\begin{equation}
  m_e\frac{d^2r}{d\xi^2} + \frac{4\pi n_0e^2}{c^2\beta_b^2} ~ \frac{1}{2r}(r^2 - r_{\rm tube}^2) = 0
 \label{eq:EoM}
\end{equation}
where $m_e$ is the electron mass. Defining the plasma frequency $\omega_p = \sqrt{\tfrac{4\pi n_0 e^2}{m_e}}$ and $\rho = r/r_{\rm tube}$, equation \eqref{eq:EoMNorm} is rewritten as
\begin{equation}
\frac{d^2\rho}{d\xi^2} + \frac{1}{2\beta_b^2} \Big( \frac{\omega_p}{c} \Big)^2 \frac{1}{\rho}(\rho^2 - 1) = 0
\label{eq:EoMNorm}
\end{equation}
The above equation is a non-linear second order differential equation and describes the natural oscillations of a plasma electron about the channel wall. The lack of charge within the channel wall gives rise to an asymmetry in these oscillations. Setting $r_{\rm tube}=0$ returns the standard simple harmonic oscillations seen in homogeneous plasmas about a cylindrical axis.

\subsection{Weakly driven surface charge dynamics: \\ 
linear surface electron oscillations}
\label{sec:LinearTheory}
Equation \eqref{eq:EoM} is readily solvable when considering small displacements of the electron from the tube wall. These small displacements are valid for very low driving beam charges or large tube radii, when the electrostatic force of the beam acting on the plasma electrons is small. In this limit, $r\approx r_{\rm tube}$, and eq.\ref{eq:EoMNorm} is linearized using
\begin{equation}
r^2-r_{\rm tube}^2 = (r-r_{\rm tube})(r+r_{\rm tube}) \approx 2r ~ (r-r_{\rm tube})
\label{eq:Approx}
\end{equation}
where the first term on the right hand side is a second order term and has been removed. The linearised equation of motion is thus
\begin{equation}
  \frac{d^2r}{d\xi^2} = - \frac{(\omega_p/c)^2}{\beta_b^2} (r-r_{\rm tube})
\label{eq:EoMNormLinear}
\end{equation}
which has the solution
\begin{equation}
  r(\xi) = r_{\rm tube} + A\sin{\left( \frac{\omega_p}{c\beta_b} \xi \right)}
\end{equation}
where $A$ is a constant. For an ultrarelativistic driving beam, $\beta_b=1$, and so an immediate form for the oscillation wavelength in the linear/weakly excited case is
\begin{equation}
  \lambda_{\rm linear} = 2\pi\frac{c}{\omega_p}
\label{eq:wlLinear}
\end{equation}
which is the well-known result for plasma oscillations in homogeneous plasma.

\subsection{Strongly driven surface charge dynamics:\\ non-linear surface oscillations}
\label{sec:LinearTheory}
Solving equation \eqref{eq:EoM} in general requires calculation of the plasma electron's initial effective velocity $\rho '$ for a given radial position $\rho$. Three basic assumptions are made to simplify the calculation to a good approximation. 

The first is that the driving beam is assumed to be a quasi-static point charge of total charge $Q_b$. This assumption is valid provided the drive beam density changes over multiple electron oscillations and its charge is conserved. Gauss' law states that the electric field intersecting a Gaussian surface $S$ is the same regardless of the shape of the charge distribution within $S$. Quasistaticity ensures the shape or size of the beam do not change significantly over time such that beam-plasma intersections do not arise.

The second assumption requires that electrons excited by the driving beam are no longer influenced by the driving beam beyond the first collapse to the axis. In other words, the primary electron collapse occurs at $\xi \gg 0$, corresponding an electric potential of approximately zero. This assumption simplifies calculation of the kinetic energy gained by the electron due to the driving beam, as the radial position of the electron at collapse need no longer be determined.

The final assumption is that the kinetic energy gained by the electron is primarily radial kinetic energy. This simplifies determination of the electron velocity at the crystal tube wall (section \ref{sec:initialVel}).

\vspace{-3.5mm}
\subsubsection{Transforming Surface oscillation equation to first order}
As equation \eqref{eq:EoM} is an autonomous ODE (i.e. an ODE with no dependence on $\xi$), the following manipulation can be made:
\begin{equation}\label{eq:AutoManip}
  \frac{d}{d\xi} \left[ \frac{1}{2} \left( \frac{d\rho}{d\xi} \right)^2 \right] = \frac{d\rho}{d\xi} \frac{d^2\rho}{d\xi^2}
\end{equation}
Using the chain rule on the left hand side of \eqref{eq:AutoManip}
\begin{equation}\label{eq:ChainManip}
  \frac{d}{d\xi} = \frac{d\rho}{d\xi} \frac{d}{d\rho}\nonumber \\
  \rho'' = \frac{d}{d\rho} \left( \frac{1}{2}\rho'^2 \right)
\end{equation}
where $\rho' = d\rho/d\xi$. Equation \eqref{eq:ChainManip} can then be substituted into \eqref{eq:EoM} and integrated, resulting in
\begin{equation}\label{eq:NewEoM}
\frac{1}{2}\rho'^2 + \frac{1}{2\beta_b^2} \left( \frac{\omega_p}{c} \right)^2 \left( \frac{1}{2}\rho^2 - \ln\rho \right) = C_1
\end{equation}
where $C_1$ is a constant of integration to be determined.

\vspace{-3.5mm}
\subsubsection{Surface oscillation: Initial Condition for Velocity}
\label{sec:initialVel}
To find $C_1$, one must know a value of $\rho' $ for a given $\rho $. At $\xi = 0$, an electron at the surface ($\rho = 1$) sees the repulsive potential (attractive potential) of the electron (positron) beam. As the electron is pulled to the axis, its energy will be converted from potential energy between it and the beam to potential energy from the no longer neutral plasma. As it recoils back towards the crystal tube wall, the electron gains kinetic energy which will become maximized at $\rho = 1$ as, beyond $\rho = 1$, the force will be directed anti-parallel to the electron velocity. Therefore, the kinetic energy of the electron at the crystal tube wall (after the excitation from the beam) is effectively equal to the potential energy it has at $(\rho, \xi) = (1,0)$ due to the electron or positron beam under the assumptions given at the start of this section.

By letting $\rho'(\rho = 1) = \rho_0'$, i.e. some initial effective velocity to be determined later, one arrives at an equation for $C_1$ after substitution into \eqref{eq:NewEoM}
\begin{align}\label{eq:C1}
C_1 = \rho_0'^2 + \frac{1}{2\beta_b^2} \left( \frac{\omega_p}{c} \right)^2
\end{align}

Determining $\rho_0'$ requires consideration of the energy gained by the electron using the simplifying assumptions made in the introduction of this section. The potential energy of an electron due to an electron or positron beam is
\begin{align}\label{eq:PotEnergy}
U(\rho,\xi) = -\frac{e^2 N_b}{4\pi\epsilon_0} \frac{1}{\sqrt{\rho^2r_{\rm tube}^2 + \xi^2}}
\end{align}
where $N_b = Q/e$ is the total number of electrons or positrons in the beam. Due to the conservative nature of the potential, the kinetic energy $E$ gained by a surface electron due to the electron or positron beam, initially at $\xi = 0$, is
\begin{align}\label{eq:potDiff}
E = U(\rho_1, \xi_1) - U(1,0)
\end{align}
where $(\rho_1, \xi_1) $ defines the position of the particle when its radial velocity is zero. If it is assumed that $\xi_1 $ is large, then $U(\rho_1, \xi_1)\approx 0$ and equation \eqref{eq:potDiff} reduces to
\begin{align}\label{eq:EnergyEq}
  E = \frac{e^2 N_b}{4\pi\epsilon_0 r_{\rm tube}} \approx \frac{1}{2}m_e\dot{r}_0^2
\end{align}
where $\dot{r}_0 = r_{\rm tube}v_b \rho_0'$ is the initial condition velocity, $v_b$ is the beam velocity, and the final term on the right hand side assumes that the energy gain occurs primarily in the radial direction. Rearranging equation \eqref{eq:EnergyEq} yields
\begin{align}
  \dot{r}_0^2 &\approx \frac{e^2 n_b}{m_e \epsilon_0} \frac{\mathcal{V}_b}{4\pi r_{\rm tube}}\nonumber\\
  &= \omega_{pb}^2 \frac{\mathcal{V}_b}{4\pi r_{\rm tube}}
\end{align}
where $N_b = n_b \mathcal{V}_b$, $\mathcal{V}_b$ is the effective volume of the beam, and $n_b$ is the beam density. For a Gaussian beam distribution of width $\sigma_r$ and length $\sigma_z$, $n_b$ is defined as the electron or positron density at the beam's centre (or the peak density) with $\mathcal{V}_b = \sigma_r^2 \sigma_z \sqrt{2\pi}$. Upon substituting $\dot{r_0} = r_{\rm tube}v_b\rho_0'$ into the above equation and rearranging, an approximate form for $\rho_0'$ is determined as
\begin{align}
\rho_0' \approx \frac{\omega_{pb}}{c} \frac{1}{\beta_br_{\rm tube}^{3/2}} \sqrt{\frac{\mathcal{V}_b}{2\pi}}
\end{align}
and, for a Gaussian beam profile
\begin{align}
\rho_0' \approx \frac{\omega_{pb}}{c} \frac{1}{\beta_br_{\rm tube}^{3/2}} \sqrt{\frac{\sigma_r^2 \sigma_z}{\sqrt{2\pi}}}
\end{align}

\vspace{-3.5mm}
\subsubsection{Non-linearity parameter and radial boundary conditions}\label{sec:alphaparam}
Due to the oscillatory nature of the problem, it is clear that there will exist two solutions for $\rho $ where $\rho' = 0$. After substituting expressions for $C_1$ and $\rho_0'$, equation \eqref{eq:NewEoM} reduces to
\begin{align}\label{eq:TranEq}
  \rho^2 - 2\ln\rho &= 1+2\,\frac{n_b}{n_0}\frac{\mathcal{V}_b}{2\pi r_{\rm tube}^3}
\end{align}
Defining $\alpha = 1+2\, \tfrac{n_b}{n_0} \tfrac{\mathcal{V}_b}{2\pi r_{\rm tube}^3} $, equation \eqref{eq:TranEq} describes a transcendental equation with two solutions:
\begin{align}
  \rho_+ &= \sqrt{-W_{-1}(-e^{-\alpha})} \\
  \rho_- &= \sqrt{-W_0(-e^{-\alpha})} \\
  \alpha &= 1+2\,\frac{n_b}{n_0}\frac{\mathcal{V}_b}{2\pi r_{\rm tube}^3} \\
 \Big( &= 1+2\,\frac{n_b}{n_0}\frac{\sigma_r^2 \sigma_z}{\sqrt{2\pi} r_{\rm tube}^3}, \,\textrm{Gaussian Profile} \Big)
\end{align}
where $W_{-1, 0}(x)$ are the decreasing and increasing branches of the lambert $W$ function respectively. Each solution respectively describes the amplitude of the crests and troughs of the plasma density oscillations. Looking at the extreme cases for $\alpha $

\begin{subequations}
\begin{align}
\rho_+ &\rightarrow  \begin{cases}
  1 & \alpha \rightarrow 1 \\
  \infty & \alpha \rightarrow \infty
  \end{cases}\\
\rho_- &\rightarrow  \begin{cases}
  1 & \alpha \rightarrow 1 \\
  0 & \alpha \rightarrow \infty
  \end{cases}
\end{align}
\end{subequations}
which suggests that, for large plasma densities and tube wall radii or physically small, low density beams, $|\rho_+ -1| \approx |\rho_- - 1|$ yielding a linear wave. In the opposite case, the wave amplitudes are different and thus the wave is non-linear. It is therefore deduced that $\alpha$ must describe the strength of non-linearity of the wave, and that increasing $n_b$, $\sigma_r$, $\sigma_z$, or decreasing $n_0$ or $r_{\rm tube}$ results in increased non-linearity.

Increasing the beam charge will correspond to a stronger driving potential experienced by the plasma electrons. As a consequence, electrons have more energy to collapse closer to the axis. This is similar for the crystal tube radius; electrons will initially be closer to the driving beam and thus experience a stronger potential. Conversely, decreasing the plasma density for a fixed beam density will reduce the number of plasma ions which weakens the restoring force allowing the tube wall electrons to collapse closer to the axis.

\vspace{-3.5mm}
\subsubsection{Wavelength of surface density oscillation: analytical model}
\label{sec:Wavelength}
Equation \eqref{eq:NewEoM} can be rearranged in terms of $\rho'$, leading to an integral solution $\xi(\rho) $ with no closed form expression:
\begin{align}\label{eq:xiSoln}
\xi(\rho) - \xi_0 = \frac{c\beta_b}{\omega_p} \int_1^\rho \frac{dx}{\sqrt{\tfrac{1}{2}\alpha - \left(\tfrac{1}{2}x^2 - \ln x \right)}}
\end{align}
where $\xi_0$ is a constant of integration and $x$ is a dummy variable.

Equation \eqref{eq:xiSoln} is restricted to a range spanning half the wavelength of the density oscillation, and a domain of $(\rho_-, \rho_+)$. The wavelength is thus
\begin{subequations}
\begin{align}
  \lambda &= 2 \big[ \xi\big( \rho_+(\alpha) \big) - \xi \big( \rho_-(\alpha) \big) \big]\\
          &= 2\frac{c\beta_b}{\omega_p} \int_{\rho_-(\alpha)}^{\rho_+(\alpha)} \frac{dx}{\sqrt{\tfrac{1}{2}\alpha - \left(\tfrac{1}{2}x^2 - \ln x \right)}} =2\frac{c}{\omega_p}I(\alpha) \label{eq:intSoln}
\end{align}
\end{subequations}
where in the last expression \ref{eq:intSoln}, $\beta_b=1$. The integral $I(\alpha)$ converges to $\pi$ as $\alpha\rightarrow 1$, which is consistent with the linear solution. However, the wavelength of nonlinear surface oscillations is greater than that in the linear regime by the factor $2 \times I(\alpha)$,
\begin{align}
  \lambda_{\rm crunch-in} = 2 \times I(\alpha) ~ 2\pi c{\omega_p}^{-1}
\end{align}
The $\omega_p$ dependence in \eqref{eq:intSoln} is as per expectations that the density oscillation wavelength is strongly affected by the plasma density in the tube walls.

The parameter $\alpha$ also affects the wavelength in the model, suggesting a dependence of the wavelength on changing $n_b$, $\sigma_r$, $\sigma_z$ and $r_{\rm tube}$. Thus $\alpha$ the nonlinearity factor provides a correction to the oscillation wavelength under linear approximation. The factor $I(\alpha)$ does not include the effect of relativistic enlargement of the wavelength of radial oscillations which is a well-known additional factor.

This model therefore suggests the wavelength of oscillation may be tuneable by adjusting the plasma density $n_0$, keeping $\alpha$ constant. Conversely, the model suggests the possibility of directly controlling the strength of non-linearity while maintaining a constant wavelength, simply by adjusting multiple parameters at once.

The dependence on the plasma density, $n_0$, outside of the integral agrees with simulation data that adjusting the plasma frequency results in strong changes in wavelength. In addition, the solution suggests that the wavelength is tunable by adjusting $n_0$ while $\alpha $ remains constant. Conversely, the model suggests the possibility of directly controlling the strength of non-linearity while maintaining a constant wavelength, simply by adjusting multiple parameters at once.

\FloatBarrier
\section{Proof-of-Principle simulations results:\\
solid-state electron oscillations with Particle-In-Cell simulations}
\label{sec:crunch-in-PIC-simulations}

Multi-dimensional Particle-In-Cell simulations using the EPOCH code \cite{epoch-pic} have been carried out to model collective electron oscillation phenomena in solid-state or crystal plasma. The use of a PIC code for analyzing collective electron oscillations at densities, $n_0 > \rm 10^{22} cm^{-3}$ is justified under the conditions where the phenomena is collision-less as was shown above to be the case under strong excitation of valence electrons in crystals. Moreover, as the length of the driver particle beam is chosen to be of the order of the wavelength of collective electron oscillations, it is possible to sustain plasmonic oscillations of electrons without triggering phonons and other mixed modes etc. Furthermore, crystal tube structures are known to naturally have mean free path lengths of several hundreds of nanometers \cite{nanotubes-Ijima}.

In this work we model the interaction of a crystal tube with intense sub-micron scale electron beam, especially its bunch length being sub-micron scale. We restrict the choice of maximum crystal tube internal diameter to 1 micron. Currently, crystal tubes are grown by folding multiple layers of mono-atomic sheets into a cylinder and in the process closing a sheet upon itself. The precise nano-engineering process of growing hundreds of nanometer tube radius is yet to be fully characterized to determine the electron density profile spanning the cross-section from the edge of the tube wall to the axis of the tube. 

Graphene based carbon nanotubes (CNT) are in recent years shown to be relatively straightforward to manufacture in the sense that sophisticated machinery is generally not required \cite{nanotubes-Ijima}. These crystal tubes have valence electron densities in the range of $10^{22-24} cm^{-3}$ and a mean free path of about $\sim$ 1 micron. These tube characteristics are ideally suited for supporting collective electron oscillations. Commercially, silica ($\rm SiO_2$) or molten glass based tubes with internal diameters ranging from 200 - 1000nm are sold variously as nano-capillary \cite{molex-catalog} etc. The gradient of density at the interface of the tube wall and hole region has been characterized using scanning electron microscope (SEM) and found to be less than its nanometer scale resolution limit. Similarly, the uniformity of the diameter is found to be quite consistent over several meters of tube length. However, the extent of surface deformities and imperfections of the exact tubes grown from atomic monolayer deposition is not precisely modeled or characterized here and will form a part of the future work.

Our computational modeling effort characterizes an experimentally realizable interaction scenario due to the recently reported (planned) availability of sub-micron scale bunch lengths at Stanford Linear Accelerator Center \cite{yakimenko-facet-ii} and possibly at other accelerator facilities in the near future.

This computational modeling effort currently serves as a proof-of-principle and is an attempt to demonstrate the possibilities that can be opened up by the solid-state tube wakefield acceleration technique. In our modeling we utilize the fact that beams of several hundred nanometer bunch lengths are accessible. For experimental relevance the modeling effort is carried out under the following constraints:
\begin{enumerate}[label=(\roman*),topsep=3pt, itemsep=0.3ex,partopsep=0.3ex, parsep=0.3ex]
\item In sec.\ref{sec:beam-waist-lt-tube-radius} - an experimentally available beam density of $n_b = 1.0 \times 10^{21} \rm cm^{-3}$ is considered. The beam waist-size is $\sigma_r = 500nm$ which is also chosen to be the tube radius $r_{\rm tube} = 500nm$. The PIC simulations indicate that a tube wall electron density of $n_{\rm tube} = 2.0 \times 10^{22} \rm cm^{-3}$ is suitable. This tube wall density may need customized nanoengineering.
\item In sec.\ref{sec:beam-waist-gt-tube-radius} - under an experimental constraint that the beam waist-size exceeds the tube diameter we model a beam-tube interaction scenario such that the beam particles radially in the Gaussian wings of the tube interact with the tube wall, $\sigma_r > r_{\rm tube}$ while the most intense part of the beam propagates within the tube. 
\item In sec.\ref{sec:n-tube-1e23-n-beam-1e23} - we assume that a beam density of $n_b = 1.0 - 5.0 \times 10^{22} \rm cm^{-3}$ is experimentally accessible. In this case the suitable tube wall densities of $n_{\rm tube} = 1.0 - 5.0 \times 10^{23} \rm cm^{-3}$ are known to be available commercially off-the-shelf.
\end{enumerate}

\vspace{-3.5mm}
\subsection{Beam waist comparable with tube radius, $r_{tube}\gtrsim\sigma_r$}
\label{sec:beam-waist-lt-tube-radius}

Using PIC simulations we model and make preliminary investigations of beam-driven solid-state tube acceleration in a parameter regime where the waist-size of the beam injected into a crystal tube is comparable to the crystal tube radius.

The electron density in the tube walls is chosen to be $n_0 = \rm 2.0 \times 10^{22} cm^{-3}$ with a fixed ion background. In the simulation results presented below a 2D cartesian grid is chosen such that it resolves the reduced plasmonic wavelength of $\rm\lambda_{pe}/(2\pi) = 38nm$ with 15 cells in the longitudinal and 15 cells in the transverse direction. Thus each grid cell in these simulations is about 2.5nm x 2.5nm (the Debye length, conservatively assuming a few eV thermal energy is $\rm \lambda_D \simeq 1\AA$). The cartesian box co-propagates with the electron beam. The box dimensions span $\rm 7\mu m$ in longitudinal direction and at least $\rm 7\mu m$ in the transverse (it is wider in transverse to incorporate wider beams). The tube electrons are modeled with 10 particle per cell of the cartesian grid. Absorbing boundary conditions are used for both fields and particles.

The electron beam has a $\gamma_b = 10,000$ (roughly 5.1 GeV) with a Gaussian bunch profile of a fixed bunch length with $\sigma_z = 400nm$. Typical beam density of $\rm 1.0 \times 10^{21} cm^{-3}$ is considered in this experimentally relevant modeling effort. The beam is initialized with 16 particle per cell. In order to analyze the interaction, the waist-size of the beam, $\sigma_r$ and the tube radius $r_{\rm tube}$ are varied. The moving simulation box tracks the particle beam. The particle beam is initialized in vacuum and propagates into the crystal tube before the simulations box begins to move.

\begin{figure}[!htb]
\vspace{-3.0mm}
\centering
   \includegraphics[width=0.6\columnwidth]{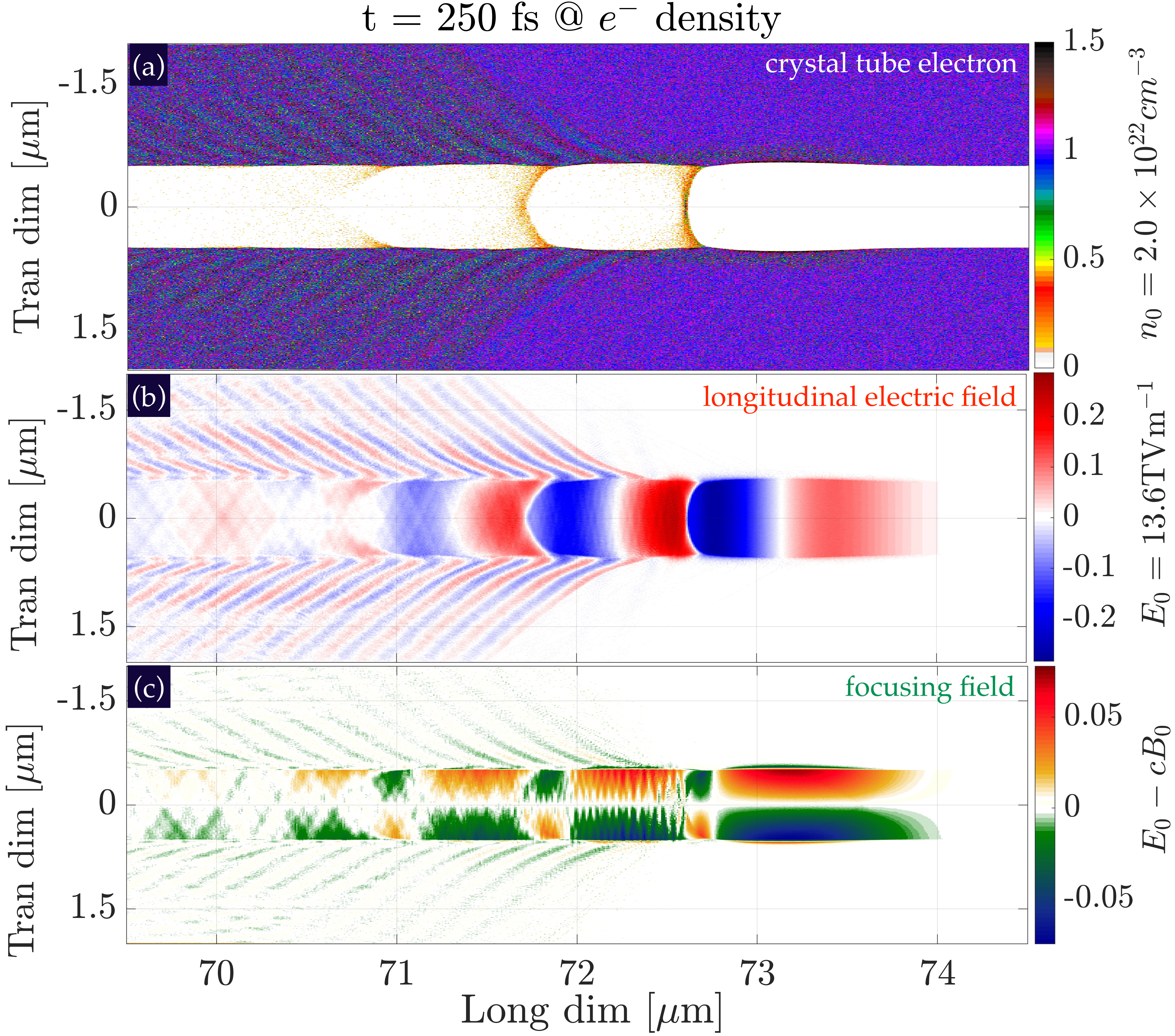}
   \vspace{-2.0mm}
   \caption{2.5D PIC simulation snapshot at around $72\mu m$ of beam-tube interaction showing solid-state tube accelerator with crystal tube radius of 500nm and beam waist-size, $\sigma_r = 500nm$ and bunch length, $\sigma_z=400nm$. The beam density is $n_b=\rm1.0\times10^{21}cm^{-3}$ whereas the channel wall density of the tube is $n_{\rm tube} = \rm 2.0\times10^{22}cm^{-3}$. Solid-state tube wakefield accelerator dynamics extracted from a 2.5D PIC simulation at around $72\mu m$ of beam-tube interaction showing the tube wall electron density (in a, normalized to $n_{\rm tube} = \rm 2.0\times10^{22}cm^{-3}$), longitudinal electric field of the surface wave (in b, normalized to $E_0=\rm13.6~TVm^{-1}$) and the focusing field (in c, $E_0 - cB_0$).}
\label{fig:wakefields-waist-lte-tube}
\vspace{-3.5mm}
\end{figure}

PIC simulation snapshots in Fig.\ref{fig:wakefields-waist-lte-tube} correspond with solid-state tube accelerator interaction parameters of crystal tube radius, $\rm r_{tube}$ of 500nm and beam waist-size, $\sigma_r = 500nm$ and bunch length, $\sigma_z=400nm$. From this snapshot we observe that a surface wave is sustained by the oscillations of the electrons across the interface of the tube wall with density, $n_{\rm tube} = \rm 2.0\times10^{22}cm^{-3}$. In Fig.\ref{fig:wakefields-waist-lte-tube}(a) the tube wall density snapshot in real-space shows {\it three} distinct spatial oscillations of a surface plasmon sustained by radial electron oscillations across the surface. These snapshots are at a simulation time of 250 fs which corresponds to a beam-tube interaction length of around $72\mu m$ (the interaction has a delayed start as the beam is initialized in vacuum and pushed to propagate into the plasma).

Beam density profiles from PIC simulation for the drive beam with $\sigma_z=400nm$ at a density of $n_b=\rm1.0\times10^{21}cm^{-3}$ show that the beam electrons experience the transverse or focusing fields of the ``crunch-in'' wakefields of the surface wave and exhibit betatron oscillations. This effect of beam density modulation can be prominently observed in the beam density snapshots presented in later sections in Fig.\ref{fig:e-beam-density-tube-waist-gt-tube}(b) and Fig.\ref{fig:HD_tube_beam_densities_n03e23_nb5e22}(d). These coherent density modulations of the drive beam were first modeled in an innovative plasma beam dump proposal \cite{Wu-plasma-dump-2010}. This coherent drive beam density modulation has been observed and labelled as scalloping in some recent works. In the near-term, beam-tube interaction can be experimentally studied by observing the small spatial-scale drive beam density modulations.

In Fig.\ref{fig:wakefields-waist-lte-tube}, the longitudinal (in b) and focusing forces (in c) are shown along with the density wave in real-space. The fields are normalized to the Tajima-Dawson acceleration gradient ($E_{wb}=E_0=m_ec\omega_{pe}e^{-1}$ or the cold-plasma wavebreaking limit). The Tajima-Dawson limit for the tube density of $n_{\rm tube} = \rm 2.0\times10^{22}cm^{-3}$ is $E_0 = \rm 13.6TVm^{-1}$. 

The simulation results lead to several interesting possibilities. It is observed that although the beam to tube density ratio $n_b/n_{\rm tube}$ is only 0.05, the longitudinal wakefields approach $\langle E_{\rm acc} \rangle \simeq 0.25 E_0$ which is an acceleration gradient of around $\langle E_{\rm acc} \rangle \simeq \rm 2.5TVm^{-1}$. Moreover, the ``crunch-in'' behavior results in the excitation of focusing fields of the order of 0.1 $E_0$ which is a focusing gradient of several $\rm 100 GVm^{-1}$.

\begin{figure}[!htb]
\vspace{-3.0mm}
\centering
   \includegraphics[width=0.8\columnwidth]{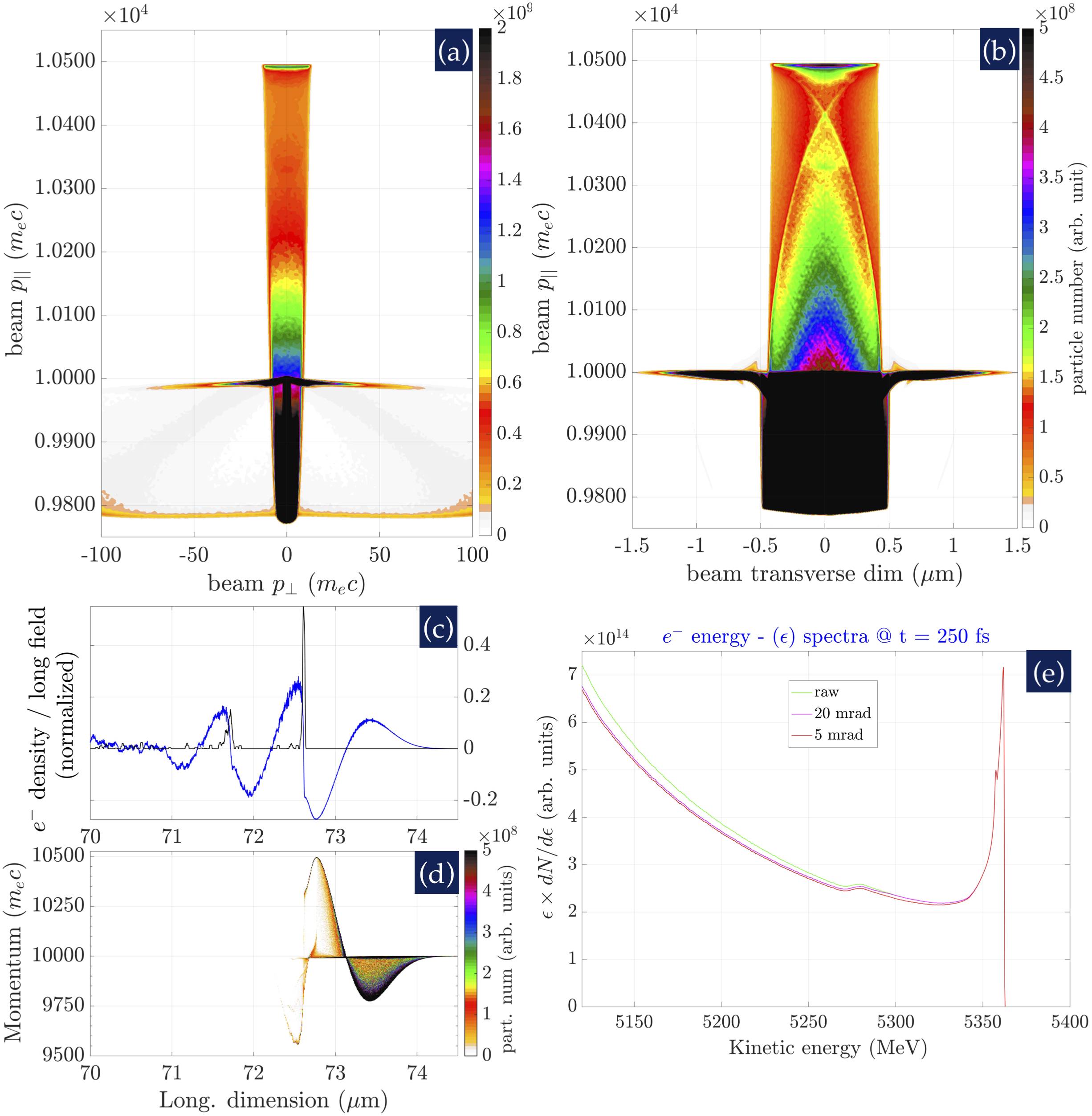}
   \vspace{-2.0mm}
   \caption{Longitudinal momentum phase-space against transverse momentum (in a) and transverse real space dimension (in b) from 2.5D PIC simulation snapshot after around $72\mu m$ of beam-tube interaction with parameters same as Fig.\ref{fig:wakefields-waist-lte-tube}. Longitudinal phase-space against longitudinal real space dimension (in c) and corresponding on-axis field, electron density lineout (in d) from 2.5D PIC simulation snapshot are also shown around $72\mu m$ of beam-tube interaction. }
\label{fig:long-phase-space-waist-lte-tube}
\vspace{-2.5mm}
\end{figure}

Fig.\ref{fig:long-phase-space-waist-lte-tube} describes the longitudinal phase-space of beam-tube interaction at around $72\mu m$ in the ``crunch-in'' surface wakefields regime. From the longitudinal momentum against transverse momentum is shown in Fig.\ref{fig:long-phase-space-waist-lte-tube}(a) and transverse real space dimension in Fig.\ref{fig:long-phase-space-waist-lte-tube}(b) it can be observed that only those beam particles that are within the crystal tube and that experience both the longitudinal as well as the transverse ``crunch-in'' wakefields undergo acceleration. Longitudinal phase-space plotted against longitudinal real space dimension in Fig.\ref{fig:long-phase-space-waist-lte-tube}(c) and the corresponding on-axis field, electron density lineout in Fig.\ref{fig:long-phase-space-waist-lte-tube}(d) demonstrate that the longitudinal dimension of the wakefield is such that the beam particles in the tail of the beam undergo acceleration. This opens up the possibility of using a single hundreds of nanometer scale bunch to observe a few $\rm TVm^{-1}$ acceleration gradients.

The energy spectra in Fig.\ref{fig:long-phase-space-waist-lte-tube}(e) shows some of the particles in the tail of the beam being accelerated from the initial beam energy centered around 5110MeV to 5360MeV, a gain of about 250MeV in $72\mu m$. This gives an average acceleration gradient over $72\mu m$ of around $\langle E_{\rm acc} \rangle \simeq \rm 3.0 TVm^{-1}$. It is also quite evident from the snapshots in (a) and (b) that only those beam electrons that are within the tube get accelerated in the surface wave, whereas the electrons in the Gaussian wings of the beam undergo minimal energy change.

The possibilities of accessing high average acceleration gradients of the order of several $\rm TVm^{-1}$ are unprecedented and being based upon a modeling effort where realistic parameters are utilized call out for an experimental verification campaign.

\begin{figure}[!htb]
\vspace{-3.0mm}
   \includegraphics[width=\columnwidth]{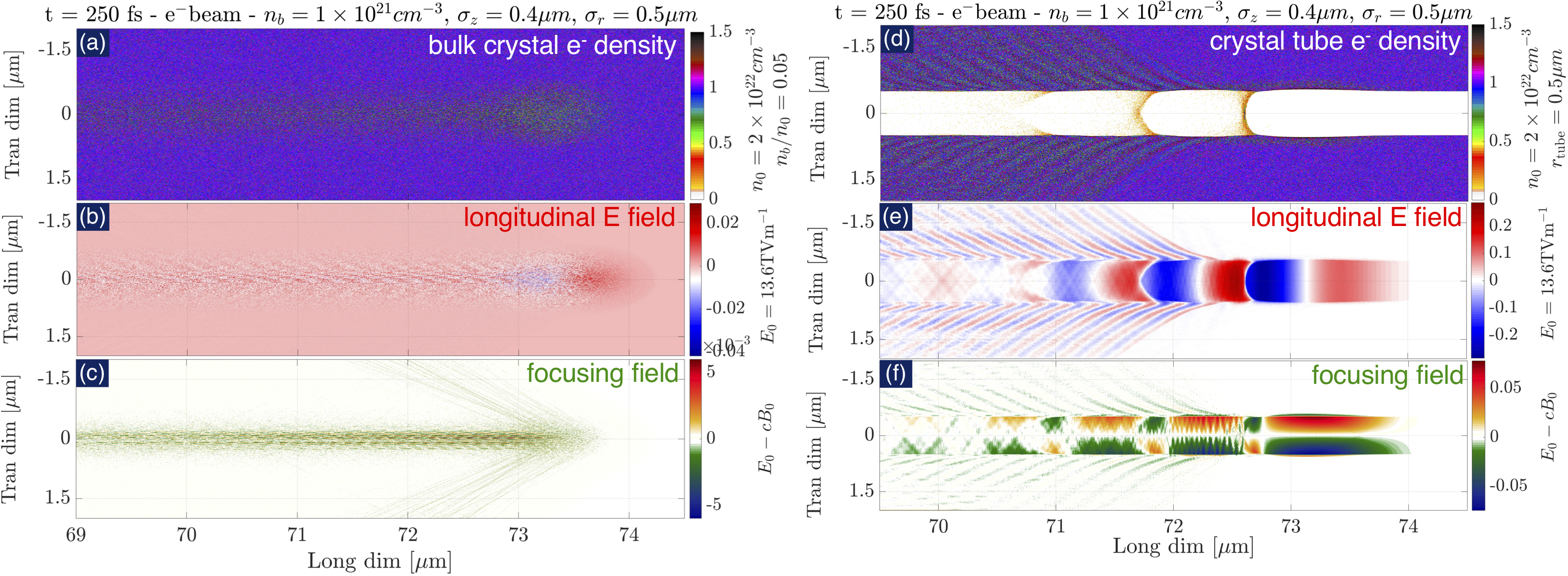}
   \vspace{-5.0mm}
   \caption{Comparison of homogeneous plasma wakefield (in a,b,c) with the crystal tube wakefield (in d,e,f; repeated from Fig.\ref{fig:wakefields-waist-lte-tube}) from 2.5D PIC simulation snapshot after around $72\mu m$ of beam-tube interaction with parameters same as Fig.\ref{fig:wakefields-waist-lte-tube}. The comparison of (a) and (d) shows that whereas the surface wave wakefield for $n_b/n_0=0.05$ is in the nonlinear regime, the wakefields driven in homogeneous plasma are almost non-existent due to the low beam to plasma density ratio.}
\label{fig:comp-bulk-vs-tube-wakefield}
\vspace{-3.0mm}
\end{figure}

\FloatBarrier
In Fig.\ref{fig:comp-bulk-vs-tube-wakefield}, a comparison of bulk plasma ($n_0 = n_{tube} = 2 \times 10^{22} cm^{-3}$) and crystal tube wakefields is presented by plotting side-by-side the electron density (in a,d), longitudinal electric field (in b,e) and the focusing field (in c,f) profiles.  It is quite evident from the comparison of the snapshots in (a) and (d) that whereas the surface wave wakefield amplitude for $n_b/n_0=0.05$ is significantly high and in the nonlinear regime for a tube radius, $r_{\rm tube} = 500nm$, the wakefields driven in homogeneous plasma are almost non-existent due to the low beam to plasma density ratio.

\begin{figure}[!htb]
\vspace{-3.0mm}
\centering
   \includegraphics[width=0.6\columnwidth]{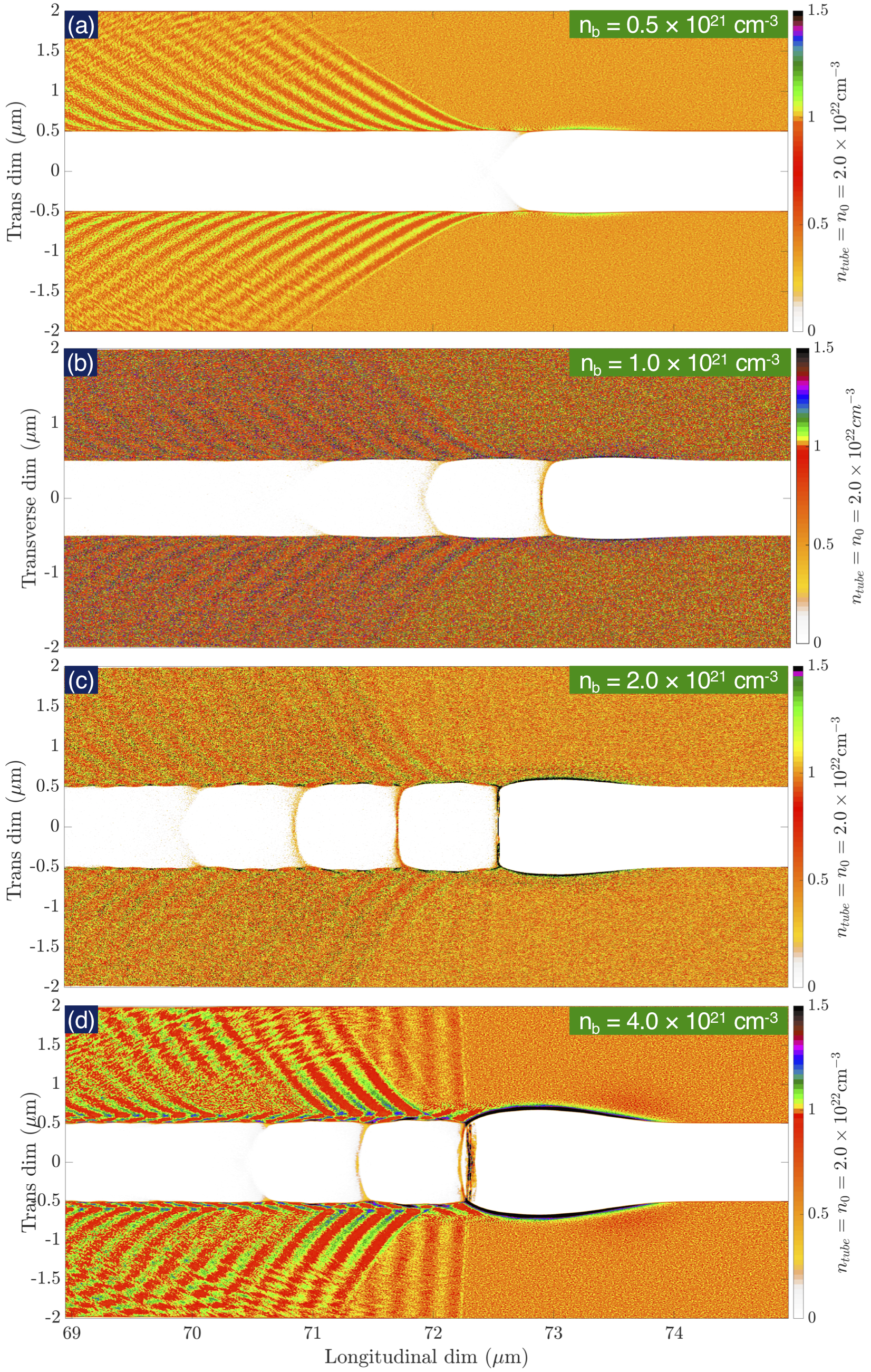}
   \vspace{-2.0mm}
   \caption{Comparison of the ``crunch-in'' surface wave modes in crystal nanotube for different drive beam densities from 2.5D PIC simulation snapshot after around $72\mu m$ of beam-tube interaction in a crystal tube with wall density, $n_{\rm tube} = \rm 2.0\times10^{22}cm^{-3}$. The drive beam densities are $n_b=$: (a) $0.025~n_{tube}$ (b) $0.05~n_{tube}$ (c) $0.1~n_{tube}$ (d) $0.2~n_{tube}$.}
\label{fig:increasing-nonlinearity-tube-density-waist-lte-tube}
\vspace{-3.0mm}
\end{figure}

\FloatBarrier
In order to characterize the effect of the ratio of drive beam density to tube wall density, $n_b / n_{\rm tube}$, in Fig.\ref{fig:increasing-nonlinearity-tube-density-waist-lte-tube} we compare the PIC simulation snapshots of tube electron density for different $n_b / n_{\rm tube}$ ratio at a beam-tube interaction length of $72\mu m$. These beam densities are: (a) $n_b=0.5\times10^{21}cm^{-3} = 0.025~n_{tube}$ (b) $n_b=1.0\times10^{21}cm^{-3} = 0.05~n_{tube}$ (c) $n_b=2.0\times10^{21}cm^{-3} = 0.1~n_{tube}$ (d) $n_b=4.0\times10^{21}cm^{-3} = 0.2~n_{tube}$.

The corresponding peak longitudinal on-axis fields or acceleration gradient over varying beam density as extracted from PIC simulations are summarized as follows:
\begin{enumerate}[label=(\alph*),topsep=3pt, itemsep=0.3ex,partopsep=0.3ex, parsep=0.3ex]
\item $\langle E_{\rm acc} \rangle \simeq 0.1 E_0$ for $n_b=0.5\times10^{21}cm^{-3} = 0.025~n_{tube}$
\item $\langle E_{\rm acc} \rangle \simeq 0.25 E_0$ for  $n_b=1.0\times10^{21}cm^{-3} = 0.05~n_{tube}$
\item $\langle E_{\rm acc} \rangle \simeq 0.65 E_0$ for $n_b=2.0\times10^{21}cm^{-3} = 0.1~n_{tube}$
\item $\langle E_{\rm acc} \rangle \simeq 1.5 E_0$ for $n_b=4.0\times10^{21}cm^{-3} = 0.2~n_{tube}$
\end{enumerate}


It is quite evident that as the drive beam density is increased in ``crunch-in'' regime of solid-state tube, the surface electron trajectories become increasing nonlinear. The nonlinear surface wave results in wakefields that are not only higher (of the order of the Tajima-Dawson acceleration gradient limit) but also result in the excitation of stronger focusing fields within the tube.


\vspace{-3.5mm}
\subsection{Beam waist size larger than the tube diameter, $\sigma_r>r_{tube}$}
\label{sec:beam-waist-gt-tube-radius}

An experimentally accessible parameter regime in the short-term where the beam waist-size is a few times larger than the crystal tube radius is investigated below using preliminary PIC simulations. In these simulations it is assumed that the peak of the beam density coincides with the axis of the crystal tube such that the most intense part of the externally focussed beam travels in the low density region of the tube. The PIC simulation setup and beam density are as described in sec.\ref{sec:beam-waist-lt-tube-radius}. 

From the PIC simulation results that are summarized below in Fig.\ref{fig:e-beam-density-tube-waist-gt-tube}, it is quite clear that surface wave wakefields in the ``crunch-in'' regime are sustained within the tube even when $\sigma_r > r_{tube}$. From the results in this section we observe that when the beam density is retained same, spatial profiles of the wakefields and the acceleration gradient of the order of $\rm 2.5 TVm^{-1}$ sustained in the case of $\sigma_r>r_{tube}$ being studied here are nearly equal to the case where $\sigma_r \leq r_{tube}$.

This excitation of near Tajima-Dawson acceleration gradient (0.1$E_{wb}$) limit in a crystal tube is quite interesting because the peak drive beam density of $\rm 1.0\times 10^{21} cm^{-3}$ is much smaller than the tube wall density of $\rm 2.0\times 10^{22} cm^{-3}$. The simulations show an energy gain of about 50MeV in around $23.5\mu m$ which is an average acceleration gradient of $\langle E_{\rm acc} \rangle > \rm 2.0 TVm^{-1}$. 

Moreover, the ``crunch-in'' regime wakefields observed in this work show that strong coherent focusing fields are also excited within the tube of the order of several $\rm 100 GVm^{-1}$. It is evident from the beam density snapshots in Fig.\ref{fig:e-beam-density-tube-waist-gt-tube} that only those beam particles that are within the tube and which as a result experience the $\langle E_{\rm acc} \rangle \sim \rm TVm^{-1}$-scale fields of the surface plasmon wave undergo significant density perturbation.

From comparison of the density and wakefield characteristics of crystal tube wakefields in Fig.\ref{fig:wakefields-waist-lte-tube} for the case of $\sigma_r \sim r_{\rm tube}$ and the same (not shown) for the case of $\sigma_r \gg r_{\rm tube}$, it is observed that the wakefield characteristics, amplitude and spatial profile, are quite similar. As the drive electron beam density in the case of Fig.\ref{fig:wakefields-waist-lte-tube} and $\sigma_r \gg r_{\rm tube}$ case are equal, $n_b = 1.0 \times 10^{21}  \rm cm^{-3}$, it is possible to postulate that drive beams of a given density are equally effective at the excitation of crystal tube wakefields irrespective of their transverse properties (given that the peak of the beam spatial distribution is aligned with the axis of the tube).


\begin{figure}[!htb]
\vspace{-3.0mm}
\centering
   \includegraphics[width=0.85\columnwidth]{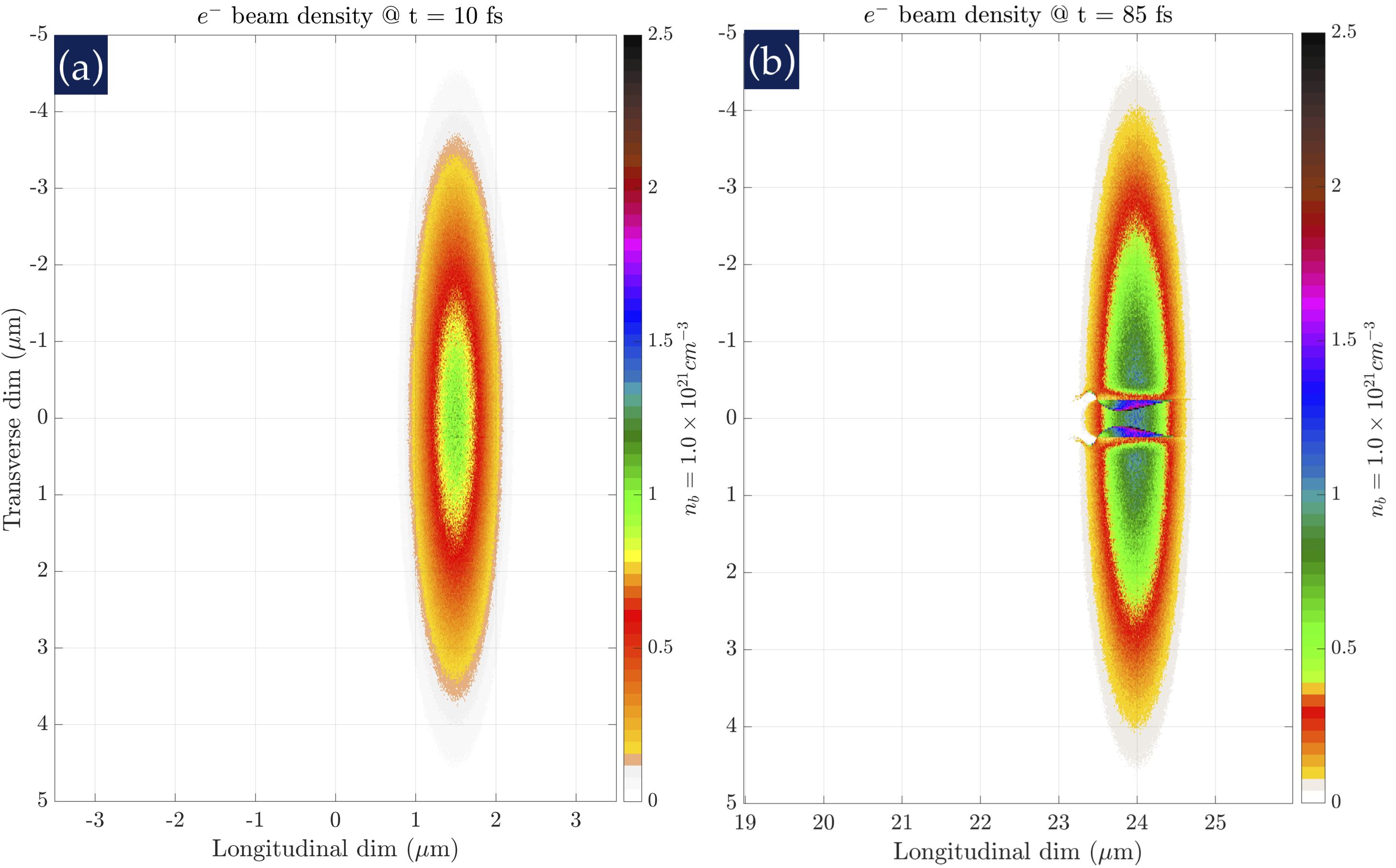}
   \vspace{-2.0mm}
   \caption{2.5D PIC simulation snapshots comparing the electron beam density at initialization and after around $23.5\mu m$ of beam-tube interaction.}
   \label{fig:e-beam-density-tube-waist-gt-tube}
\vspace{-3.5mm}
\end{figure}

\FloatBarrier
\subsection{Scaling to off-the-shelf tube wall densities: $n_{\rm tube} \sim 10^{23} \rm cm^{-3}$}
\label{sec:n-tube-1e23-n-beam-1e23}

Crystal nanotubes \cite{nanotubes-Ijima} that are currently available off-the-shelf have mass densities in the range of $1.3-2.0 ~\rm g\text{-}cm^{-3}$. For purely Carbon atom based nanomaterial this mass density translates to ionic and electron densities in highly ionized states of between $n_{\rm tube} \sim 10^{23-24} \rm cm^{-3}$. In this section we present our examination of the scaling of the ``crunch-in'' modes in crystal tubes when the crystal wall densities are around $n_{\rm tube} \sim 10^{23-24} \rm cm^{-3}$. The PIC simulation setup and beam density are the same as described in sec.\ref{sec:beam-waist-lt-tube-radius}. These simulations show that if certain beam densities may be experimentally within reach of existing electron beam facilities, then it may be possible to excite beam-driven solid-state tube surface wave wakefields using off-the-shelf nanotubes.

In the previous sections, sec.\ref{sec:beam-waist-lt-tube-radius} and sec.\ref{sec:beam-waist-gt-tube-radius}, we have presented proof-of-principle PIC simulation results under the constraint that the beam densities are limited to around $n_b \leq 2 \times 10^{21} \rm cm^{-3}$ whereas the beam bunch length is characterized by $\sigma_z \simeq 400 \rm nm$. Under this constraint on the beam density, the ``crunch-in'' regime was observed to be accessible only for tube wall densities, $n_{\rm tube} \sim 10 \times n_{b}$.

\begin{figure}[!htb]
\vspace{-3.0mm}
\centering
   \includegraphics[width=0.9\columnwidth]{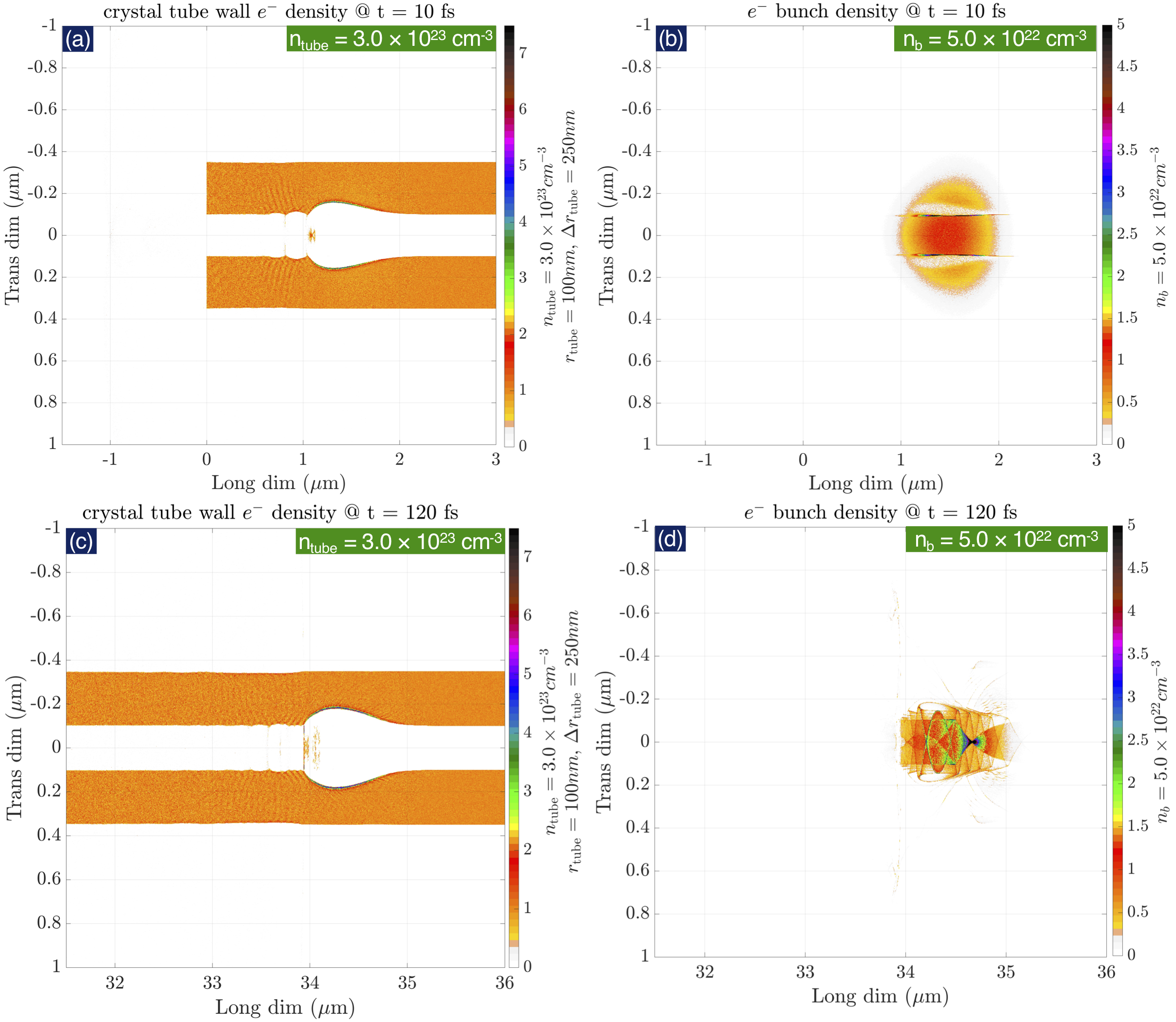}
   \vspace{-2.0mm}
   \caption{Density snapshots for off-the-shelf solid-state tube parameters from 2.5D PIC simulations showing tube electron density (in a,b) with fixed background ions and drive beam electron density (in b,d) after around $36\mu m$ of beam-tube interaction. From (a,c) it follows that ``crunch-in'' mode is excited and that the beam evolves as it experiences the transverse fields of this mode.}
   \label{fig:HD_tube_beam_densities_n03e23_nb5e22}
\vspace{-3.5mm}
\end{figure}

In this section, we assume that beam densities as high as $n_b \sim 5.0 \times 10^{22} \rm cm^{-3}$ are experimentally accessible using currently available accelerator facilities. With these range of beam densities, using PIC simulations snapshots presented below, we observe that coherent ``crunch-in'' wakefields supported by collective oscillations of crystal electrons are accessible using nanotube structures that are available off-the-shelf. Therefore, if beam densities of the order of $n_b \sim 1.0 \times 10^{23} \rm cm^{-3}$ are experimentally accessible, then proof of concept experimental verifications of SOTWA mechanism can be carried out in the near term.

In the simulations presented in this section the electron density in the tube walls is chosen to be $n_0 = \rm 3.0 \times 10^{22} cm^{-3}$ with a fixed ion background. A 2D cartesian grid is chosen such that it resolves the reduced plasmonic wavelength of $\rm\lambda_{pe}/(2\pi) = 10nm$ with 20 cells in the longitudinal and 20 cells in the transverse direction. Thus each grid cell in these simulations is about $\rm 500\AA$ x $\rm 500\AA$ (the Debye length is $\rm \lambda_D \leq 1\AA$). 

\begin{figure}[!htb]
\vspace{-3.0mm}
\centering
   \includegraphics[width=0.45\columnwidth]{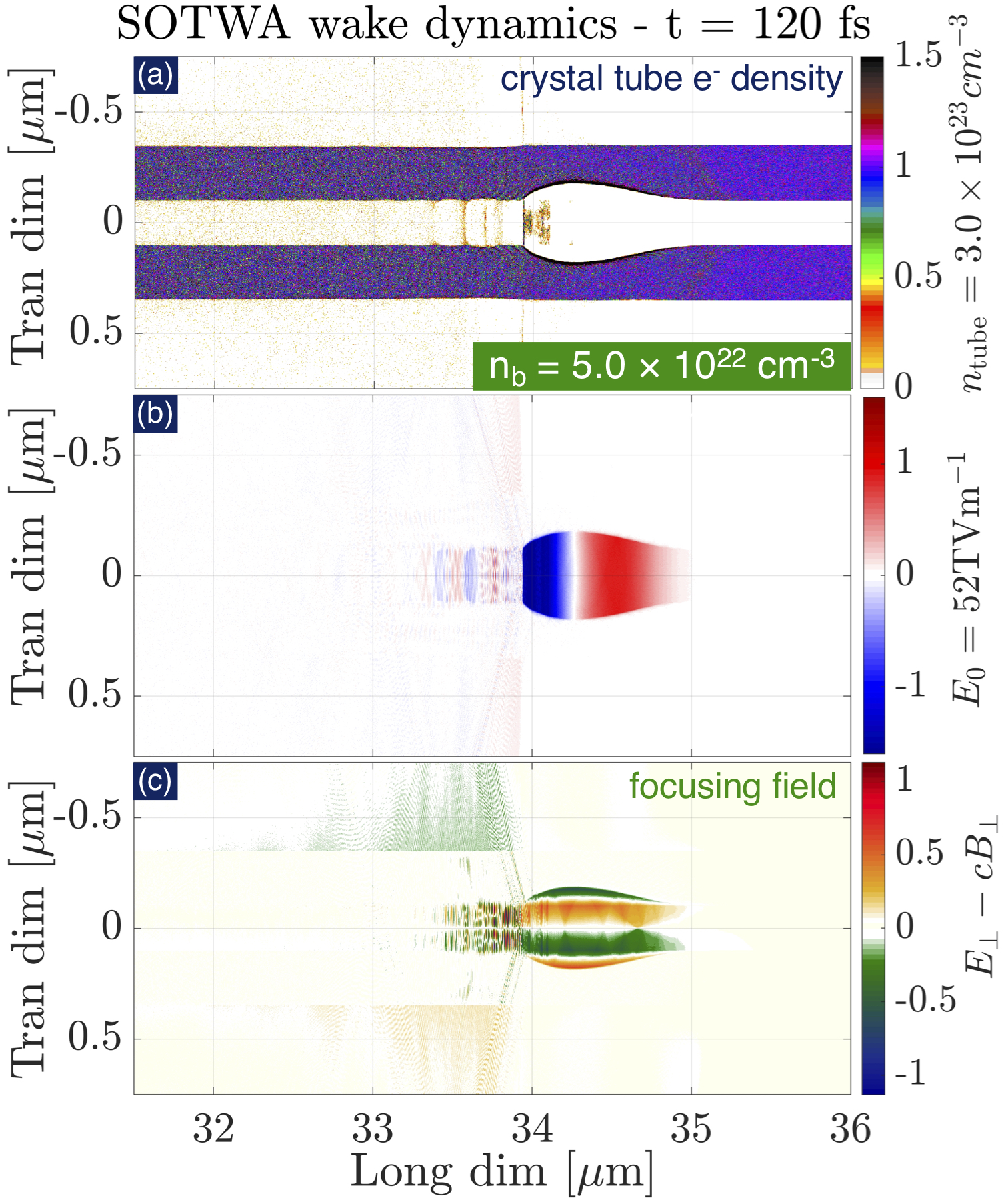}
   \vspace{-2.0mm}
   \caption{2.5D PIC simulation snapshot of the tube wall electron density (in a), longitudinal field (in b) and focusing field (in c) at around $36\mu m$ of beam-tube interaction with exactly the same beam and tube parameters as in Fig.\ref{fig:HD_tube_beam_densities_n03e23_nb5e22}. From (b,c) both the longitudinal as well as transverse wakefields are of the order of the Tajima-Dawson acceleration gradient limit, $52\rm TVm^{-1}$.}
   \label{fig:HD_dynamics_n03e23_nb5e22}
\vspace{-2.0mm}
\end{figure}

The crystal tube of radius, $r_{\rm tube} = 100nm$ is here is modeled to have a finite thickness of 250nm, with the outer radial extent of the tube thus terminating at 350nm from the axis of the tube. The cartesian box co-propagates with the electron beam. The box dimensions span $\rm 5\mu m$ in longitudinal direction and at least $\rm 3\mu m$ in the transverse. The tube electrons are modeled with 4 particle per cell.

The electron beam has $\gamma_b = 10,000$ (roughly 5.1 GeV) with a Gaussian bunch profile of a fixed bunch length with $\sigma_z = 400nm$. In the simulation snapshots presented below the beam density is $\rm 5.0 \times 10^{22} cm^{-3}$ which is experimentally relevant. The beam is initialized with 9 particle per cell. A comparison of the PIC simulation results in sec.\ref{sec:beam-waist-lt-tube-radius} and sec.\ref{sec:beam-waist-gt-tube-radius} provides enough confidence that the critical parameter in beam tube interaction is the beam density, $n_b$ with the $\sigma_r$ to $r_{\rm tube}$ ratio being relatively insignificant. In consideration of this we use a beam with $\sigma_r = 250nm$, with a good and previously justified approximation that a beam of higher waist-size (for example, $\sigma_r \sim 2.5 \mu m$) but the same density will have the same characteristics of beam tube interactions and excite considerably similar wakefields.

From these simulation snapshots summarized in Fig.\ref{fig:HD_tube_beam_densities_n03e23_nb5e22},\ref{fig:HD_dynamics_n03e23_nb5e22} it is possible to conclude that if beam densities as high as $n_b \sim 5.0 \times 10^{22} \rm cm^{-3}$ are experimentally accessible at current accelerator facilities, then it may be possible to excite strong ``crunch-in'' wakefields in off-the-shelf crystal tubes of nominal tube dimensions. In our simulations, the tube has a radius ($r_{\rm tube}$) of 100nm and a wall thickness ($\Delta r_{\rm tube}$) of 250nm. The beam density is initialized to $n_b = 5.0 \times 10^{22} \rm cm^{-3}$ and tube wall density is initialized to $n_{\rm tube} = 3.0 \times 10^{23}  \rm cm^{-3}$, with the $n_b/n_{\rm tube} = 0.17$. The beam properties are: $\gamma_b = 10,000$, $\sigma_z = 400nm$ and $\sigma_r = 250nm$.

\begin{figure}[!htb]
\vspace{-3.0mm}
\centering
   \includegraphics[width=0.85\columnwidth]{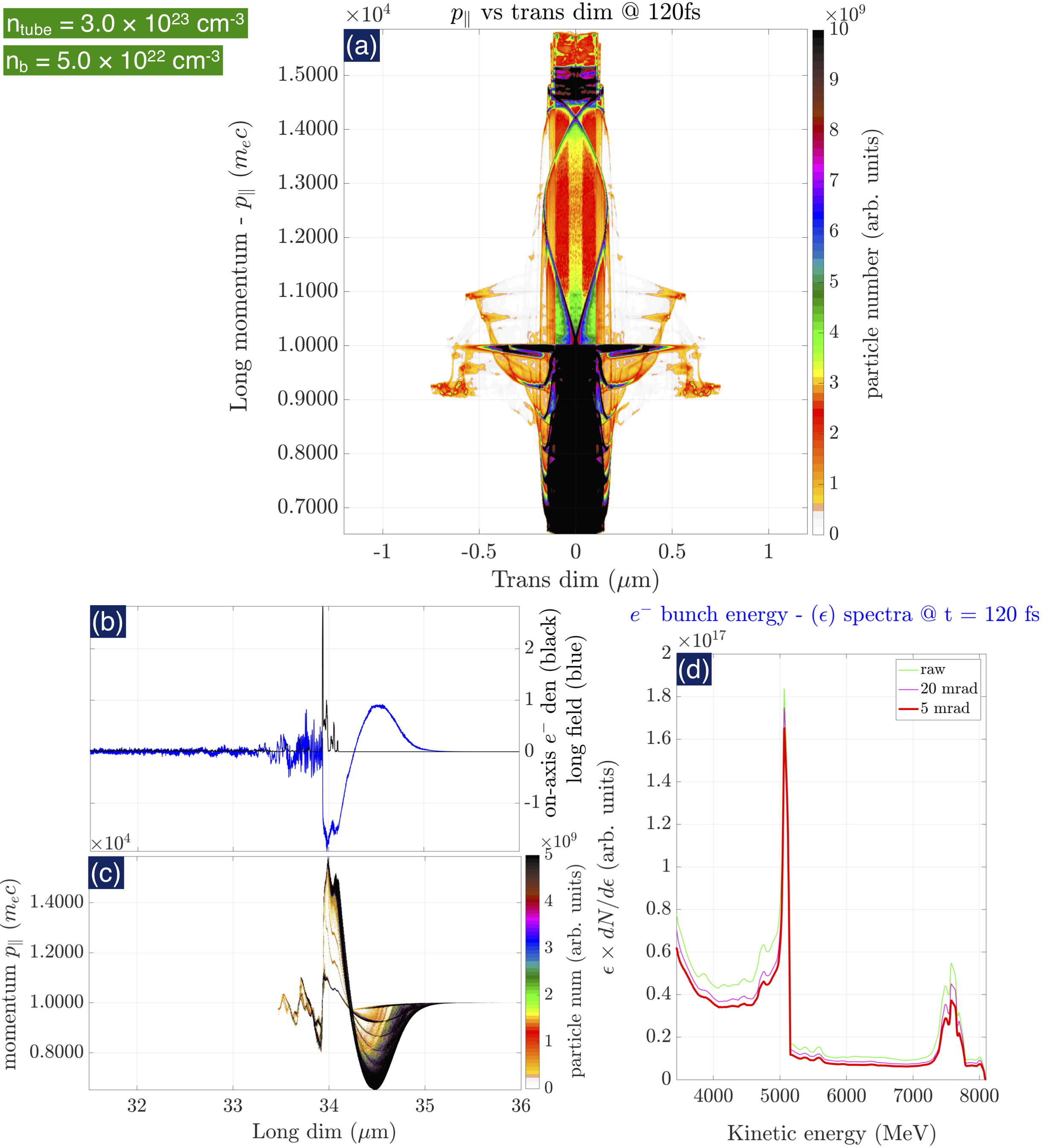}
   \vspace{-2.5mm}
   \caption{Longitudinal momentum phase-space against transverse real space dimension (in a), the same against longitudinal real space dimension (in c) and the corresponding on-axis field, electron density lineout (in b) from 2.5D PIC simulation snapshot after $36\mu m$ of beam-tube interaction.}
   \label{fig:HD_long_phase_space_n03e23_nb5e22}
\vspace{-2.5mm}
\end{figure}

From \ref{fig:HD_long_phase_space_n03e23_nb5e22} we can infer that a few $10 \rm ~ TVm^{-1}$ acceleration gradient may be experimentally realizable using current accelerator facilities. The accelerated energy spectra shown in \ref{fig:HD_long_phase_space_n03e23_nb5e22}(d) shows the acceleration of a small fraction of the drive beam from the initial beam energy centered around 5110 MeV to 7570 MeV, a gain of about 2.46GeV in $36\mu m$ gives an average gradient of around $\langle E_{\rm acc} \rangle \simeq \rm 70.0 ~ TVm^{-1}$. It is quite evident from the transverse real-space vs longitudinal momentum snapshot in (a) that only those beam electrons that are within the tube ($r_{\rm tube}=100nm$) get accelerated in the wakefield. This increase in the acceleration gradient simply follows the electron density scaling of the Tajima-Dawson acceleration gradient. 

It is quite attractive to have the possibility to accelerate a part of the 5GeV particle beam by around 2.5GeV in sub millimeter-scale crystal tubes while sustaining unprecedented many tens of $\rm TVm^{-1}$ acceleration gradients.

\section{Discussion and Future Work}

In this work we have presented a preliminary analytical and computational model of beam-driven solid-state acceleration mechanism in crystal tubes. The solid-state tube wakefield acceleration or SOTWA mechanism presented here utilizes collective electron oscillation modes on and across the surface of a crystal tube. These plasmonic oscillations sustain propagating surface waves driven as the wakefield of a charged particle beam of submicron bunch length (and, ideally submicron waist-size). A tube shaped nanostructured crystal is not only found to offer the possibility to minimize the direct high-intensity interaction of the beam with bulk crystal but also the possibility of excitation of significantly higher wakefield amplitude compared to the direct interaction of same density particle beam with bulk crystal.

The experimentally available submicron scale bunch length (for instance, planned $\sigma_z = 400nm$\cite{yakimenko-facet-ii}) is shown to have the potential for resonant excitation of collective electron oscillations in crystal tube. The resonant excitation of a surface mode in a crystal tube driven by the beam at a given density is shown to be experimentally realizable within a range of tube wall densities. A preliminary analytical model of the crystal tube wall surface electron oscillations has been presented based upon the seminal works on modeling many body crystal phenomena as collective modes of collisionless Fermi electron gas. Our model currently assumes minimal ion motion over the relevant attosecond timescales (from our preliminary mobile ion simulations) but crystal lattice ion motion effects will be a major part of the future work.

In the computational models presented in sec.\ref{sec:crunch-in-PIC-simulations} using beam densities of the order of $n_b \sim 10^{21} \rm cm^{-3}$, as some of the electrons in the tail of the drive bunch experience strong wakefields of the tube surface wave, they are shown to rapidly gain energy. The average acceleration gradients experienced by the tail particles are shown to be of the order of several $\rm TVm^{-1}$ as per the expectations of the Tajima-Dawson acceleration gradient limit for crystals. The particles in the tail of the beam gain several hundred MeVs in a few hundred microns under the influence of surface wakefields (sec.\ref{sec:beam-waist-lt-tube-radius} and sec.\ref{sec:beam-waist-gt-tube-radius}). The possibility of accessing average acceleration gradients that are at least {\it two} orders of magnitude higher than the gaseous plasma wakefield acceleration techniques will pave the way forward in accelerator research. 

It is further demonstrated that if experimentally accessible beam densities may be of the order of $n_b \sim 5.0 \times 10^{22} \rm cm^{-3}$ with other beam properties being the same, then off-the-shelf crystal tubes of a few hundred nanometer diameter can be utilized (sec.\ref{sec:n-tube-1e23-n-beam-1e23}). With the densities of the these off-the-shelf tubes being of the order of $10^{23} \rm cm^{-3}$, the accessible Tajima-Dawson acceleration gradients are of the order of $\rm 10 ~ TVm^{-1}$. Our simulations suggest the possibility of SOTWA fields being at least {\it three} orders of magnitude higher than gaseous plasma acceleration technology.

The possibility that an increase in the drive beam density allows access to a nonlinear surface wave ``crunch-in'' regime has been demonstrated. In the ``crunch-in'' regime both strong transverse fields of the order of many $\rm 100GVm^{-1}$ as well as longitudinal wakefields of the order of many $\rm TVm^{-1}$ are excited. Thus, this regime using a crystal tube is useful to control the accelerated bunch transverse properties while the accelerated particles do not directly experience high ion density in their propagation path resulting in the minimization of associated instabilities. 

{\bf Controlled crystal tube photon source:} Moreover, the strong transverse fields of the crystal tube wake make controlled and tunable generation of gamma-ray photons as an electron or positron beam particle trajectories undergo oscillations during their interaction with ``crunch-in''  transverse fields of the order of many $\rm 100GVm^{-1}$. The use of specifically structured crystal tube such as with a superlattice, allows significantly higher control of the gamma-ray flux as opposed to the uncontrolled filamentation driven interaction in a metal \cite{beam-crystal-filamentation}.

{\bf Nano-modulation of drive beam:} In the very near-term, beam-tube interaction can be experimentally investigated by the observation of coherent density modulations of the drive beam \cite{Wu-plasma-dump-2010} which may be related to the effect of beam scalloping observed in some gaseous plasma studies {elec-Beam-Wakefield-Expt}. Beam-tube interaction can be experimentally diagnosed by observing the small spatial-scale beam density modulations after the interaction.

In future work, we will extend the analytical and computational modeling of the SOTWA mechanism presented here. Moreover, we will determine the optimal conditions for the excitation of strong many $\rm TVm^{-1}$ acceleration (and in-tube focusing) gradients under various tube and beam parameters by modeling the plasmonic surface wave in crystal tube. It is also critical to understand non-ideal conditions of the interaction, such as misalignment of the beam and tube axis, effect of limited thickness of tube wall and electron density profile of tube, secondary high-field ionization of the channel walls, ion motion processes, modification of tube density profile due to ablation of the crystal tube to beam irradiation etc. The extent and time-scales of damage caused to the crystal tube structure by the drive beam, the possibility of reuse in consideration of effects such as atomic stabilization as well as the effect of these non-ideal structural properties on collective electron oscillations will also be carefully modeled. 

{\bf Laser Wakefield Accelerator injector for SOTWA:} Our future work will also study the external injection of the inherently micron-scale electron and positron beams \cite{Sahai-positron-LPA} that are accelerated using laser-driven wakefields in gaseous plasmas and are thus likely to be more accessible. Furthermore, it is well known that the radiation from muons interacting with the high transverse or focusing field of the tube and undergoing oscillations is significantly smaller than electrons or positrons ($\propto (m_e/m_{\mu})^4$), we will also model the injection and acceleration of muons. Laser acceleration of muons \cite{Sahai-muon-LPA}, also put forth and investigated as part of this XTALS 2019 workshop, is modeled to be able to produce ultra-short micron-scale muon beams that are suitable for injection into crystal tube wakefields.

Through the proposed extensive modeling effort, our work will seek the parameter regime and feasible diagnostics for demonstration of an experimental prototype of possibly many $\rm TVm^{-1}$ average acceleration gradient of the SOTWA mechanism.

\section{Acknowledgment}
A. A. S. was supported by the College of Engineering and Applied Science, University of Colorado, Denver. 
V. D. S. was supported by Fermi National Accelerator Laboratory, which is operated by the Fermi Research Alliance, LLC under Contract No. DE-AC02-07CH11359 with the United States Department of Energy.
We appreciate valuable discussions with S. Chattopadhyay, U. Winenands, G. Stupakov and V. Lebedev. 
This work used the Extreme Science and Engineering Discovery Environment (XSEDE), which is supported by National Science Foundation grant number ACI-1548562 \cite{xsede-citation}. This work utilized the RMACC Summit supercomputer through the XSEDE program, which is supported by the National Science Foundation (awards ACI-1532235 and ACI-1532236), the University of Colorado Boulder, and Colorado State University \cite{rmacc-citation}.


\end{document}